\documentclass[amsfonts,amssymb,aps,eqsecnum]{revtex4}

\def\unitvec#1{\hat#1}

\begin{document}
\newtheorem{thm}{Theorem}
\title{Gleason-Type Derivations of the Quantum Probability Rule for
Generalized Measurements}
\author{Carlton M.~Caves,$^1$ Christopher
A.~Fuchs,$^{2,}$\protect\footnote{Temporary Address: Communication
Networks Research Institute, Dublin Institute of Technology,
Rathmines Road, Dublin 6, Ireland.} Kiran Manne,$^1$ and Joseph
M.~Renes$^1$} \affiliation{$^1$Department of Physics and
Astronomy, University of New Mexico,
\\
Albuquerque, New Mexico 87131--1156, USA
\\
$^2$Quantum Information and Optics Research, Bell Labs, Lucent
Technologies,
\\
600--700 Mountain Avenue, Murray Hill, New Jersey 07974, USA}
\date{2003 June~26}

\begin{abstract}
We prove a Gleason-type theorem for the quantum probability rule using
frame functions defined on positive-operator-valued measures (POVMs),
as opposed to the restricted class of orthogonal projection-valued
measures used in the original theorem. The advantage of this method is
that it works for two-dimensional quantum systems (qubits) and even for
vector spaces over rational fields---settings where the standard
theorem fails. Furthermore, unlike the method necessary for proving the
original result, the present one is rather elementary. In the case of a
qubit, we investigate similar results for frame functions defined upon
various restricted classes of POVMs. For the so-called trine
measurements, the standard quantum probability rule is again recovered.
\end{abstract}

\maketitle

\section{Introduction}
\label{sec:intro}

Two theorems position themselves at the foundation of quantum theory:
the Kochen-Specker theorem~\cite{kochenspecker67} and Gleason's
theorem~\cite{gleason57}. The former rules out the possibility of
predicting experimental outcomes with certainty, and---failing this
certainty---the latter prescribes the permitted probabilistic
descriptions.  Both are pivotal to an understanding of quantum
mechanics that begins with primitive notions about measurement---the
subject of this paper.

To adopt this perspective, imagine that for any experiment, there exist
mathematical objects which represent the possible outcomes.  Likewise,
there exists a mathematical object the experimenter uses to describe
the system under investigation.  The minimal task of any physical
theory is to determine what these objects are and to use them to
furnish the probabilities for the various outcomes.

Here the traditional viewpoint of physical theory is stood on its head.
Usually the system and its properties---whatever can be said about
them---are regarded as fundamental.  Physical systems are considered
objectively, outside the context of measurements, real or potential.
Within this traditional viewpoint the question for physics is this:
What are the properties of physical systems, and what laws do they
obey?  Notoriously this leads to the quantum measurement problem.
Having considered physical systems to obey quantum mechanics, what are
we to make of measurements?  How does measurement fit into the
framework of quantum theory?

In this paper we view measurements as fundamental, obviating the
measurement problem. The main task is an operational one: How should
measurements be described, and what predictions about their outcomes
can be made? Properties of physical systems, e.g., position and
momentum, are useful if they help in this task, but are ultimately
secondary.  Though the conventional measurement problem does not arise
in this context, a new question emerges, just as daunting as the
original: Why the Hilbert-space description of measurement
outcomes~\cite{Fuchs02,foulis95,ludwig83}?

Leaving this question aside as too ambitious for the present, we start
by assuming the form of measurements as given by classical or quantum
theory. In classical physics, the measurement objects are the points in
a phase space, while in quantum physics, they are traditionally
one-dimensional projectors on a Hilbert space. Classically only one
measurement exists---a full accounting of the phase space. In quantum
mechanics, on the other hand, any complete set of one-dimensional
orthogonal projectors suffices.

The description of the system can be given, in classical mechanics, by
a phase-space point.  This point is the ``true'' point---others are
``false''---so the outcome of a measurement can be predicted with
certainty.  Attempting such a concrete description in quantum mechanics
is ruled out by the Kochen-Specker theorem: There is no way to assign
truth and falsity to all the one-dimensional projectors in such a way
that in any measurement there is only one true outcome.  At this point
in the development, quantum mechanics becomes an irreducibly
probabilistic theory; the possibility of underlying certainties has
been ruled out.

With certainties ruled out, Gleason's theorem delineates the allowable
descriptions of the system, i.e., the form the probabilities can take.
Keeping with the linear structure, every outcome probability is an
inner product of the corresponding measurement projector and a {\em
density operator\/} for the system. The density operator---any convex
combination of one-dimensional projection operators---represents the
description or ``state'' of the system.  Thus Gleason's theorem gives a
means to go from the structure of measurements to the structure of
states. It immediately implies the Kochen-Specker result, as there are
no density operators that yield probability distributions for all
measurements that are valued only on zero and one.

Neither theorem holds for two-dimensional quantum systems, so-called
qubits, as long as the measurement objects are restricted to being
one-dimensional projection operators.  We can include this outstanding
case, however, by widening the class of allowed measurements to include
positive-operator-valued measures (POVMs), comprised of measurement
operators called {\em effects}, and in so doing, we also considerably
simplify the derivation. This generalization is the subject of this
paper. Section~\ref{sec:effects} describes how measurement outcomes in
quantum mechanics are associated with effects.
Section~\ref{sec:Gleason} shows that given the structure of effects,
the usual quantum-mechanical probability rule follows simply, even for
qubits.  Section~\ref{sec:qubits} investigates several specific
restricted classes of measurements for qubits and what kinds of
probability distributions these classes permit.

It should be emphasized from the outset that this is an inherently {\em
noncontextual\/} approach, meaning that a shared outcome of two
distinct measurements has the same probability in both contexts. This
fact is not a consequence of this approach, but rather an assumption
used in its construction.  In this construction a description of the
measurement device and a description of the physical system are
composed into a probability for each possible outcome.  In defining
measurements by their physical set-ups, this approach initially
pertains to each measurement situation individually.  The Hilbert-space
formalism we are using to describe measurements takes a further step by
associating the same measurement object with outcomes in different
measurements.  For this reason, we {\em assume\/} that these outcomes
have the same probabilities, this being the noncontextual assumption.
To abandon this assumption at the level of finding allowed probability
assignments via Gleason's theorem would be to ignore, wholly or
partially, the framework provided by the vector-space structure of
measurements.  There are other, perhaps stronger arguments for
noncontextuality~\cite{Fuchs02}, but the argument here is a minimal one
that all the authors of this paper are willing to accept.

\section{Effect operators}
\label{sec:effects}

In the typical von Neumann formulation of quantum measurement theory,
measurements are described by complete sets of orthogonal projection
operators.  Here we consider quantum measurements in their full
generality, the so-called POVMs~\cite{kraus83,busch97}. A POVM is also
a complete set of operators resolving the identity operator, but
comprised of positive operators less than the identity.  These
operators, called {\em effects}, can also be characterized as Hermitian
operators having eigenvalues in the unit interval.  The set of effects
in $d$ dimensions we denote by $\mathcal{E}_d$.

That $\mathcal{E}_d$ is a convex set is clear.  The projection
operators of all ranks, including the zero operator $\mathbf{0}$ and
the identity operator $\mathbf{1}$, form the extreme points of
$\mathcal{E}_d$, a fact that we demonstrate via the following
construction.  For a given effect $E$, order its $d$ eigenvalues
(including zero eigenvalues) from smallest to largest,
$\{\lambda_1,\dots,\lambda_d\}$.  Associated with each eigenvalue
$\lambda_j$ is a one-dimensional projector $\pi_j$ onto the
corresponding eigenvector; these projectors can be chosen to make up a
complete, orthonormal set.  Now form the projection operators
$\Pi_{k}=\sum_{j=k}^d \pi_j.$ Clearly $\Pi_1=\mathbf{1}$ and
$\Pi_d=\pi_d$.  These projectors get smaller in rank as the index gets
bigger.  The effect $E$ can be written as the following convex
combination:
\begin{equation}
\label{eqn:effectexpansion}
E=
\lambda_1\Pi_1+
\sum_{m=2}^d(\lambda_m-\lambda_{m-1})\Pi_m+
(1-\lambda_d)\mathbf{0}\;.
\end{equation}
The zero operator is included to make the sum of the coefficients unity
without affecting $E$.

Since every effect can be expanded as a convex combination of
projectors, only projectors can be extreme points of $\mathcal{E}_d$.
To show that all the projectors are extreme points, we need to show
that a projector $\Pi$ cannot be written as a {\em proper\/} convex
combination of other projectors, i.e., cannot be written as
$\Pi=a\Pi_1+(1-a)\Pi_2$, where $0<a<1$ and $\Pi_1\ne\Pi_2$.  To show
this, suppose $\Pi$ could be so written.  For any normalized vector
$|\psi\rangle$, we have $\langle\psi|\Pi|\psi\rangle=
a\langle\psi|\Pi_1|\psi\rangle+ (1-a)\langle\psi|\Pi_2|\psi\rangle$. If
$|\psi\rangle$ is in the null subspace of $\Pi$, i.e., is orthogonal to
the support of $\Pi$, we have $\langle\psi|\Pi|\psi\rangle=0$, which
implies that
$\langle\psi|\Pi_1|\psi\rangle=\langle\psi|\Pi_2|\psi\rangle=0$; this
shows that the supports of $\Pi_1$ and $\Pi_2$ are contained in the
support of $\Pi$.  If $|\psi\rangle$ is in the support of $\Pi$, we
have $\langle\psi|\Pi|\psi\rangle=1$, which implies that
$\langle\psi|\Pi_1|\psi\rangle=\langle\psi|\Pi_2|\psi\rangle=1$; this
shows that the support of $\Pi$ is contained in the supports of $\Pi_1$
and $\Pi_2$.  Together these conclusions imply that $\Pi_1=\Pi_2=\Pi$,
contradicting our assumption of a proper convex combination for $\Pi$.
We conclude that the extreme points of $\mathcal{E}_d$ are the
projectors of all ranks, including $\mathbf{0}$.

Projection operators are a limiting case of effect operators, the
latter being a ``fuzzy'' or ``unsharp'' version of the former by convex
combination.  Similarly, the von Neumann projective measurements are a
limiting case of POVMs.

In two dimensions the set of effects, $\mathcal{E}_2$, has an appealing
geometric picture.  Beginning with the parameterization of Hermitian
operators by the Pauli matrices, we write the general two-dimensional
effect as
\begin{equation}
E=r\mathbf{1}+\vec{s}\cdot\vec{\mathbf{\sigma}}=
r\mathbf{1}+s\unitvec{n}\cdot\vec{\sigma}\;,
\label{eq:2deffects}
\end{equation}
where $\unitvec n$ is a unit vector.  The eigenvalues $r\pm
||\vec{s}||=r\pm s$ must lie in the unit interval, which is equivalent
to the conditions $0\leq r\leq 1$, $0\leq s\leq r$, and $0\leq s\leq
1-r$.  Since $s$ characterizes the radius of a sphere, the full set can
be pictured in the following way: Starting with the unit interval for
$r$, associate with each point a ball of radius $r$ for $r\leq 1/2$ and
of radius $1-r$ for $r>1/2$.  The set of two-dimensional effects is
thus the intersection of two three-dimensional cones, both having an
opening angle of $45^\circ$, one extending up from a vertex at $r=s=0$
($E=\mathbf{0}$) and the other extending down from a vertex at $r=1$,
$s=0$ ($E=\mathbf{1}$). The intersection of the boundaries of the two
cones, $r=s=1/2$, is the surface of the Bloch sphere, where the effects
are one-dimensional projectors.  The effects on the boundary of the
lower cone, $r=s\le1/2$, are multiples of one-dimensional projectors.

\section{The quantum probability rule}
\label{sec:Gleason}

The task of quantum theory is, minimally, to associate with every
measurement a probability distribution for its outcomes.  We assume
this is done noncontextually for the reasons given above.  The
probability rule is thus a function from the set of effects to the
unit interval, which is normalized on any subset that makes up a
POVM.  Such a function is known as a {\em frame function}.  More
precisely, a frame function is a function $f:\mathcal{E}_d\rightarrow
[0,1]$ that satisfies
\begin{equation}
\label{eqn:framedef}
\sum_{E_j\in X}f(E_j)=1
\end{equation}
on any subset $X=\{E_j\in\mathcal{E}_d|\sum_j E_j=\mathbf{1}\}$. This
section is devoted to proving the following Gleason-type theorem, which
was first proved by Busch~\cite{busch99}.

\begin{thm}
For every frame function $f:\mathcal{E}_d\rightarrow [0,1]$, there is
a unique unit-trace positive operator $W$ such that $f(E)=(W,E)={\rm
tr}(WE)$.
\end{thm}

\noindent The operator $W$ is the density operator that gives rise to
the frame function probabilities $f(E)$.  We divide the proof of this
theorem into several parts, each of which occupies a subsection.

\subsection{Linearity with respect to the nonnegative rationals}
\label{sec:ratlin}

Every frame function is trivially additive, for consider two POVMs,
$\{E_1,E_2,E_3\}$ and $\{E_1+E_2,E_3\}$.  Clearly both are POVMs if
either is, and the frame-function requirement immediately yields
\begin{equation}
\label{eqn:additive} f(E_1)+f(E_2)=f(E_1+E_2)\;.
\end{equation}
>From this we obtain a homogeneity property for multiplication by
rational numbers.  We can break an effect $nE$ into $m$ pieces to form
the effect $(n/m)E$.  Using the additivity property twice, we obtain
\begin{equation}
\label{eqn:rationhomog}
mf\!\left({n\over m}E\right)=f(nE)=nf(E)
\qquad\Longrightarrow\qquad
f\!\left(\frac{n}{m}E\right)=\frac{n}{m}f(E)\;.
\end{equation}
The function $f$ is thus established to be linear in the nonnegative
rationals. We can extend to full linearity by proving continuity.
Alternately, adopting the strategy of Busch~\cite{busch99}, we can
demonstrate the homogeneity of $f$, from which linearity follows
immediately.  These two arguments are taken up in turn in the next two
subsections.

\subsection{Continuity}
\label{sec:continuity}

Continuity of the frame function can be established via {\em
reductio ad absurdum\/}: A contradiction with the definition of a
frame function arises if $f$ is discontinuous.  Recall the
definition of continuity for metric spaces: $f$ is continuous at
$x_0$ if for all $\epsilon>0$, there exists a $\delta>0$ such that
$|f(x)-f(x_0)|<\epsilon$ for all $x$ satisfying $|x-x_0|<\delta$.
On the space of operators, we use the Hilbert-Schmidt inner
product $(A,B)={\rm tr}(A^\dagger B)$ and the associated norm
$|A|\equiv\sqrt{(A,A)}$.

Consider first continuity at the zero operator [since
$f(E)=f(E+\mathbf{0})= f(E)+f(\mathbf{0})$, we know that
$f(\mathbf{0})=0$].  If we assume $f$ is discontinuous at the zero
operator, then there exists an $\epsilon>0$ such that for all
$\delta>0$, there exists an effect $E$ satisfying $|E|<\delta$ and
$f(E)\ge\epsilon$.  Choose $\delta=1/N<\epsilon$, where $N$ is an
integer, and let $E$ be an effect satisfying $|E|<1/N$ and
$f(E)\ge\epsilon$.  Now $F=NE$ satisfies $|F|=N|E|<1$, which
implies that $F$ is an effect, since the sum of the squares of its
eigenvalues is less than 1.  But we also have from additivity that
$f(F)=Nf(E)\ge N\epsilon>1$, contradicting the definition of a frame
function.  We conclude that $f$ is continuous at the zero operator.

We can easily translate the continuity at $\mathbf{0}$ to the entire
set of effects.  To prove continuity at an arbitrary effect $E_0$, we
need to consider neighboring operators $E$ and the difference
$E-E_0$. Diagonalizing the difference, we can write $E-E_0=A-B$,
where $A$ is the nonnegative-eigenvalue part of the
eigendecomposition and $-B$ is the negative-eigenvalue part.  It is
clear that $A$ and $B$ are positive operators satisfying
$|A|,|B|\le|A-B|=|E-E_0|$, which implies that $A$ and $B$ are effects
(provided $|E-E_0|\le1$). Applying the frame function to the equation
$E+B=E_0+A$ yields $f(E)-f(E_0)=f(A)-f(B)$ by additivity.  Invoking
continuity at zero establishes that for every
$\epsilon=\epsilon'/2>0$, there exists a $\delta>0$ such that
$|A|,|B|<\delta\Rightarrow f(A),f(B)<\epsilon'$. Thus if
$|E-E_0|=|A-B|<\delta$, we have $|A|,|B|<\delta$ and
$|f(E)-f(E_0)|=|f(A)-f(B)|\leq|f(A)|+|f(B)|<2\epsilon'=\epsilon$.
This establishes the continuity of $f$ on all of $\mathcal{E}_d$,
which in turn shows that $f$ is a linear function on $\mathcal{E}_d$.

\subsection{Homogeneity}
\label{sec:homogeneity}

An alternative route to linearity is to prove the homogeneity of the
frame function.  Following Busch's proof, we first note that the
frame function preserves order; i.e., if $E_1<E_2$ for any pair of
measurement operators, then $f(E_1)\le f(E_2)$. This follows
immediately from the definition, for $E_1<E_2\Leftrightarrow
E_2-E_1\equiv E_3>0$, so $E_3$ is an effect. Writing $E_2=E_1+E_3$,
which implies $f(E_2)=f(E_1)+f(E_3)$ by additivity, we find that
$f(E_1)\le f(E_2)$ since $f(E_3)\ge0$.

Now the pinching theorem can be used to establish the homogeneity of
$f$.  Consider two sequences of rational numbers sharing the same
irrational limit $\alpha$: $\{q_i\}$ is an increasing sequence and
$\{p_i\}$ a decreasing sequence.  By order preservation and linearity
in the nonnegative rationals, we have
\begin{equation}
\label{eqn:irrorder}q_i f(E)
=f(q_i E)\le f(\alpha E)\le f(p_i E)=p_i f(E)
\end{equation}
for all $i$.  The pinching theorem shows that $f(\alpha E)=\alpha
f(E)$, establishing that $f$ is a homogeneous and, hence, linear
function.

\subsection{Linearity and the inner product}
\label{sec:linin}

Since the frame function is linear, it arises from an inner product. To
show this, we first extend the definition of $f$ to the entire vector
space of operators; then it is a trivial theorem of linear algebra to
recast a linear function on a vector space as an inner product.

We extend the definition of the function in the most
straightforward fashion.  Let $H$ be an arbitrary Hermitian
operator.  Every such operator can be written as the difference of
two positive operators $G_1$ and $G_2$; one way to do so is simply
to diagonalize $H$ and to let $G_1$ be the positive-eigenvalue
part and $-G_2$ the negative-eigenvalue part.  Further, for any
positive operator $G$ we can find a positive number $\alpha$ such
that $\alpha E=G$ for some effect $E$.  Now we define
$f(H)=f(G_1)-f(G_2)=\alpha_1 f(E_1)-\alpha_2 f(E_2)$.  Though the
unravelling of $H$ is not unique, the extension is.  Suppose
$H=\alpha_1 E_1-\alpha_2 E_2=\alpha_3 E_3-\alpha_4 E_4$, which
implies $\alpha_1 E_1+\alpha_4 E_4=\alpha_2 E_2+\alpha_3 E_3$.
Choose $\beta$ such that $\beta\ge\max\{\alpha_j\}$ so that we
have
\begin{equation}
\frac{\alpha_1}{\beta}E_1+\frac{\alpha_4}{\beta}E_4=
\frac{\alpha_2}{\beta}E_2+\frac{\alpha_3}{\beta}E_3\;.
\end{equation}
Since every operator is now in the original domain of $f$, we can
apply the frame function to find
\begin{equation}
\alpha_1 f(E_1)+\alpha_4 f(E_4)=\alpha_2 f(E_2)+\alpha_3 f(E_3)\;.
\end{equation}
The extension being manifestly linear, we now have a linear function
$f$ on the entire space of Hermitian operators.  It can be extended
to all operators by complexification.

To rewrite this linear function as an inner product, choose an
orthonormal operator basis $\{\tau_j\}$, and write an arbitrary
operator as $A=\sum_j \tau_j(\tau_j,A)$.  Clearly we have
$f(A)=\sum_j f(\tau_j)(\tau_j,A)$.  Now define $W$ as the unique
solution of the $d^{\,2}$ equations $f(\tau_j)=(W,\tau_j)$, so that
the frame function is $f(A)=\sum_j (W,\tau_j)(\tau_j,A)=(W,A)$.  The
$d^{\,2}$ equations are, of course, nonsingular since $\{\tau_j\}$ is
an orthonormal basis.

The nonnegativity and normalization of the frame function induce the
density operator properties of $W$, i.e., positivity and unit trace.
Given an arbitrary normalized vector $|\psi\rangle$, we have $0\le
f(|\psi\rangle\langle\psi|)=\langle\psi|W|\psi\rangle$, showing that
$W$ is positive.  The condition of unit trace follows from
normalization:
\begin{equation}
{\rm tr}\,W=(W,\mathbf{1})=\biggl(W,\sum_j E_j\biggr)
=\sum_j (W,E_j)=\sum_j f(E_j)=1\;.
\end{equation}

The reader should note that if we omit the extension of linearity to
real numbers (Secs.~\ref{sec:continuity} and \ref{sec:homogeneity}),
the arguments in Secs.~\ref{sec:ratlin} and \ref{sec:linin} demonstrate
the quantum probability rule for vector spaces over rational fields.
This result has implications for recent discussions in the literature
about the possibility of describing the finite precision of real-world
measurements via vector spaces over complex numbers with rational parts
(see, in particular,
Refs.~\cite{meyer99,clifton00,appleby02,Cabello02}). We do not dwell on
these points, noting only that our result shows that rational POVMs
cannot be assigned truth values, the only frame functions being those
derived from density operators.  In our view, given that POVMs are the
preferred description of finite-precision measurements, that is all
that needs to be said about finite-precision quantum measurements.

\section{Frame functions for qubits}
\label{sec:qubits}

The POVM version of Gleason's theorem works even for qubits, unlike
the original Gleason theorem, which was based on measurements
described by one-dimensional orthogonal projectors.  We now turn our
attention specifically to qubits and investigate whether several
restricted sets of POVMs enforce the quantum probability rule.  In
particular, we show that the quantum probability rule is necessitated
by a particular subset of POVM measurements called {\em trines}.
Other measurements are studied to shed light on what kinds of
measurements yield the quantum probability rule and why.

\subsection{General description of restricted sets of POVMs}

To start, recall that the two-dimensional effects are parameterized
by four real parameters as in Eq.~(\ref{eq:2deffects}).  In any of
the restricted sets of measurements considered in this section, the
allowed POVMS are made up of effects that are multiples of
one-dimensional projectors, i.e., $r=s\le1/2$ and $E=r(1+\unitvec
n\cdot\vec\sigma)$, and all the effects have the same value of $r$.
An allowed POVM is thus specified by a set of unit vectors that sum
to the zero vector; if there are $N$ outcomes in the POVM, then
$r=1/N$.  Finally, we assume that the allowed POVMs are rotationally
invariant; i.e., they are obtained by applying all possible
3-dimensional rotations to any particular POVM in the allowed subset.

With these assumptions, we have the following structure.  For
$N$-outcome POVMs, the allowed effects have the form
\begin{equation}
E={1\over N}(1+\unitvec n\cdot\vec\sigma)\;,
\end{equation}
where $\unitvec n$ can be any unit vector.  Frame functions defined
on this set, i.e.,
\begin{equation}
f\biggl({1\over N}(\mathbf{1}+\unitvec{n}\cdot\vec{\sigma})\biggr)
\equiv F(\hat n)=
\sum_{l=0}^\infty\sum_{m=-l}^l c_{lm}Y_{lm}(\unitvec n)\;,
\label{eq:expansion}
\end{equation}
are functions on the unit sphere and thus can be expanded in terms
of spherical harmonics.  In writing the spherical-harmonic
expansion, we are assuming that $F$ is continuous on the unit
sphere. (Please be careful to note that this extra continuity
assumption was not made in the full-fledged Gleason-type theorem
of the previous section.) Properties of the spherical harmonics
that we need in the following are the following: (i)~the
separation into $\theta$ and $\phi$ (or $n_z$ and $n_x+in_y$)
dependencies,
\begin{equation}
Y_{lm}(\unitvec n)=Y_{lm}(\theta,\phi)=
\sqrt{{(2l+1)\over4\pi}{(l-m)!\over(l+m)!}} P_l^m(\cos\theta)e^{im\phi}
=h_{lm}(n_z)(n_x+in_y)^m\;,
\label{eq:Y}
\end{equation}
where $P_l^m(x)$ is an associated Legendre function, and (ii) the changes
under conjugation, reflection, and parity,
\begin{equation}
Y_{lm}(\unitvec n)=
(-1)^mY_{l,-m}^*(\unitvec n)=
(-1)^{l+m}Y_{lm}(\pi-\theta,\phi)=
(-1)^mY_{lm}(\theta,\phi+\pi)=
(-1)^lY_{lm}(-\unitvec n)\;.
\end{equation}
Particularly useful is the form of Eq.~(\ref{eq:Y}) for $m=l$:
$Y_{ll}(\unitvec n)\propto(n_x+in_y)^l$.

The sought-after quantum rule is
\begin{equation}
\label{eqn:trinerule}
F(\unitvec n)={\rm tr}(WE)=
{1\over N}(1+\unitvec n\cdot\vec P)\;,
\end{equation}
where $\vec{P}={\rm tr}(W\vec\sigma)$ is any 3-vector such that
$||\vec{P}||\leq 1$.  The quantum rule evidently contains only
$l=0,1$ spherical harmonics, with $c_{00}=\sqrt{4\pi}/N$,
$c_{10}=\sqrt{4\pi/3}\,P_z/N$, and $c_{1,\pm1}=\sqrt{2\pi/3}\,(\mp
P_x+iP_y)/N$.

Now let the set of unit vectors $\{\unitvec n_1,\ldots,\unitvec n_N\}$
specify a ``fiducial'' POVM; the completeness property of a POVM
implies that these vectors specify a POVM if and only if
\begin{equation}
0=\sum_{j=1}^n \unitvec n_j\;.
\label{eq:constraint}
\end{equation}
Any other set, $\{R\unitvec n_j\}$, where $R$ is a three-dimensional
rotation, is also a POVM.  The frame condition is that
\begin{equation}
1=\sum_{j=1}^N F(R\unitvec n_j)=
\sum_{l,m}c_{lm}\sum_{j=1}^N Y_{lm}(R\unitvec n_j)
=\sum_{l=0}^\infty\sum_{m,r=-l}^l
c_{lm}{\cal D}^{(l)*}_{mr}(R)\sum_{j=1}^N Y_{lr}(\unitvec n_j)
\label{eq:frame}
\end{equation}
for all rotations $R$.  In writing the last equality, we use
\begin{equation}
Y_{lm}(R\unitvec n)=
\sum_{r=-l}^l Y_{lr}(\unitvec n){\cal D}^{(l)}_{rm}(R^{-1})=
\sum_{r=-l}^l Y_{lr}(\unitvec n){\cal D}^{(l)*}_{mr}(R)\;,
\label{eq:Yrotation}
\end{equation}
where ${\cal D}^{(l)}_{mr}(R)$ is the irreducible (unitary) matrix
representation of the rotation $R$ in the angular-momentum subspace
with angular momentum $l$.

We now use the fundamental orthogonality property of the irreducible
representations of the rotation group~\cite{tung93}:
\begin{equation}
\int d\mu_R\,{\cal D}^{(l)*}_{mr}(R){\cal D}^{(l')}_{m'r'}(R)
={1\over2l+1}\delta_{ll'}\delta_{mm'}\delta_{rr'}\;.
\end{equation}
Here the integration is over the invariant measure $d\mu_R$ of the
rotation group.  Noting that ${\cal D}^{(0)}_{00}(R)=1$, we can use
this orthogonality relation to invert Eq.~(\ref{eq:frame}), obtaining
the condition $c_{lm}\sum_{j=1}^N Y_{lr}(\unitvec n_j) =\delta_{l0}$
for all $l$, $m$, and $r$.  For $l=0$ this is a trivial normalization
constraint, satisfied by choosing $c_{00}=\sqrt{4\pi}/N$.  For
$l\ge1$, we can write the condition in a more illuminating,
equivalent form,
\begin{eqnarray}
&\mbox{}&\vphantom{\sum_{j=1}^N}
\mbox{$c_{lm}=0$ for $m=-l,\ldots,l$}\label{eq:frame2a}\\
l\ge1:\qquad&\mbox{}&\mbox{or}\nonumber\\
&\mbox{}&\mbox{$\displaystyle{0=\sum_{j=1}^N Y_{lr}(\unitvec n_j)}$
for $r=-l,\ldots,l$.}\label{eq:frame2b}
\end{eqnarray}
These are necessary and sufficient conditions for a frame function
$F(\unitvec n)$.

The frame conditions~(\ref{eq:frame2a}) and (\ref{eq:frame2b}) are a
potent restriction.  They say that if the $l$th harmonic is allowed in
$F(\unitvec n)$, then the unit vectors for the fiducial POVM must
satisfy the sum condition~(\ref{eq:frame2b}).  The choice of fiducial
POVM being arbitrary, the sum condition must be satisfied by the unit
vectors for all POVMS in the restricted set under consideration. This
extension from a fiducial POVM to all POVMs is, however, automatic: If
a fiducial POVM satisfies the sum condition~(\ref{eq:frame2b}), then
the rotation property~(\ref{eq:Yrotation}) of spherical harmonics
guarantees that the condition is satisfied by all POVMs in the
restricted set.

The sum condition~(\ref{eq:frame2b}) is automatically satisfied for
$l=1$ by virtue of the completeness condition~(\ref{eq:constraint}).
For higher $l$, if one finds a nonzero value of $\sum_{j=1}^N
Y_{lr}(\unitvec n_j)$ for just one value of $r$ and just one set of
POVM vectors $\{\unitvec n_j\}$, then the $l$th harmonic must be absent
from frame functions. On the other hand, if the sum
condition~(\ref{eq:frame2b}) is satisfied for a fiducial POVM, then
$F(\unitvec n)$ can contain the $l$th harmonic. The expansion
coefficients $c_{lm}$ cannot be chosen arbitrarily, of course, since
$F(\unitvec n)$ must be real and nonnegative.  Making $F(\unitvec n)$
real is trivial---simply choose $c_{l,-m}=(-1)^m c_{lm}^*$---but
delineating the region of coefficients that gives rise to nonnegative
functions is generally quite a difficult task. Nonetheless, we can
conclude that the $l$th harmonic is allowed whenever the sum condition
is met, for the following reason.  Because the spherical harmonics are
bounded functions, sufficiently small expansion coefficients $c_{lm}$
can be combined with $c_{00}=\sqrt{4\pi}/N$ without making $F$
negative.

The spherical harmonic $Y_{lr}(\hat n)$ can be regarded as the $r$th
component of a rank-$l$ spherical tensor formed from $\unitvec n$; it
is a linear combination of the components of the rank-$l$ symmetric
trace-free Cartesian tensor formed from $\unitvec n$.  The vanishing of
$\sum_{j=1}^N Y_{lr}(\unitvec n_j)$ simply says that the sum of these
tensors over all the unit vectors in a POVM vanishes. Harmonics $l=1$
and $l=2$ illustrate what is going on.  The $l=1$ spherical-tensor
components $Y_{1r}(\unitvec n)$ are linear combinations of the
Cartesian components of $\unitvec n$ and thus always sum to zero over
the unit vectors in a POVM as a consequence of the
constraint~(\ref{eq:constraint}). The $l=2$ spherical-tensor components
$Y_{2r}(\hat n)$ are linear combinations of Cartesian components of the
symmetric trace-free two-tensor $n_kn_l-{1\over3}\delta_{kl}$. Thus
$\sum_{j=1}^N Y_{2r}(\hat n_j)=0$ if and only if
\begin{equation}
\sum_{j=1}^N(n_j)_k(n_j)_l={N\over3}\delta_{kl}\;;
\end{equation}
i.e., the sum of the projectors onto the 3-vectors $\unitvec n_j$ is
proportional to the 3-dimensional identity operator.

We can make another general statement.  If the allowed POVMs are made
up of pairs of (subnormalized) orthogonal projectors---i.e.,
$-\unitvec n$ is in the set of unit vectors if $\unitvec n$ is---then
the parity property of the spherical harmonics implies that all odd
harmonics are allowed in the frame function.  This, of course, is the
reason that Gleason's theorem does not hold for qubits if the allowed
measurements are restricted to orthogonal projectors.

We turn now to applying the sum condition~(\ref{eq:frame2b}) to
particular sets of POVMs~\cite{Peresproblem}.

\subsection{Restricted sets of POVMs}

\subsubsection{Trine measurements}

Consider first the trine measurements, three-outcome measurements
described by three unit vectors equally spaced in a plane, which
thereby sum to zero.  We consider the sum $\sum_{j=1}^3
Y_{ll}(\unitvec n_j)$ for the particular trine
\begin{equation}
\unitvec n_1=\unitvec e_x\;,\qquad
\unitvec n_{2,3}=
-{1\over2}\,\unitvec e_x\pm{\sqrt3\over2}\,\unitvec e_z\;.
\end{equation}
Using $Y_{ll}(\unitvec n)\propto(n_x+in_y)^l$, we find that the sum
is proportional to $1+(-1)^l/2^{l-1}$.  This being nonzero for all
$l$ except $l=1$, we conclude that a frame function has no harmonics
higher than $l=1$. To establish the precise form of the quantum rule,
we must require that the frame function be real and nonnegative.
Having ruled out all but harmonics $l=0,1$, we can write the general
form as $F(\unitvec{n})=(1+\unitvec{n}\cdot\vec{P})/N$, where
$\vec{P}$ is a 3-vector required to be real by the reality of
$F(\unitvec n)$.  The nonnegativity of $F(\unitvec n)$ then requires
that $||\vec{P}||\leq 1$, leaving us with the standard quantum rule.

\subsubsection{Tetrahedral measurements}

Now consider the tetrahedral measurements, four-outcome measurements
whose unit vectors point to the vertices of a tetrahedron.
Tetrahedral measurements are important as the two-dimensional example
of a {\em symmetric informationally complete\/} POVM, i.e., a POVM
made up of $d^{\,2}$ rank-one effects whose pairwise inner products
are all the same.  The statistics of an informationally complete POVM
determine the density operator, and the restriction on the pairwise
inner products gives symmetric informationally complete POVMS
particularly appealing symmetry properties.

Since the vertices of a tetrahedron satisfy
\begin{equation}
\sum_{j=1}^4(n_j)_k(n_j)_l={4\over3}\delta_{kl}\;,
\label{eq:tetprop}
\end{equation}
we can conclude immediately, as discussed above, that $\sum_{j=1}^4
Y_{2r}(\unitvec n_j)=0$ for all $r$ and all tetrahedra.  This means
that a frame function for a tetrahedral measurement can contain $l=2$
harmonics and thus does not necessarily follow from the quantum
probability rule.

To investigate the possibility of higher harmonics, consider the
particular tetrahedron
\begin{equation}
\unitvec n_1=\unitvec e_x\;,\qquad
\unitvec n_2=-{1\over3}\,\unitvec e_x-{2\sqrt2\over3}\,\unitvec e_z\;,\qquad
\unitvec n_{3,4}=
-{1\over3}\,\unitvec e_x+{\sqrt2\over3}\,\unitvec e_z
\pm\sqrt{{2\over3}}\,\unitvec e_y\;.
\label{eq:tet1}
\end{equation}
For this tetrahedron the sum $\sum_{j=1}^4 Y_{ll}(\hat n_j)$ is
proportional to
\begin{equation}
1+\left(-{1\over3}\right)^l
+\left(-{1\over3}+i\sqrt{{2\over3}}\right)^l
+\left(-{1\over3}-i\sqrt{{2\over3}}\right)^l=
1+{(-1)^l+2(\sqrt7)^l\cos l\alpha\over3^l}\;,
\end{equation}
where $e^{i\alpha}=-1/\sqrt7+i\sqrt{6/7}$. It is not hard to verify
that this quantity is nonzero for all $l\neq 1,2,5$
($\cos2\alpha=-5/7$ and $\cos5\alpha=-121/49\sqrt7$), implying that a
frame function cannot have any harmonics other than $l=0,1,2,5$.

To show that a frame function can have $l=5$ harmonics, rotate
the tetrahedron~(\ref{eq:tet1}) by $-90^\circ$ about the $y$ axis,
obtaining
\begin{equation}
\unitvec n_1=\unitvec e_z\;,\qquad
\unitvec n_2=-{1\over3}\,\unitvec e_z+{2\sqrt2\over3}\,\unitvec e_x\;,\qquad
\unitvec n_{3,4}=
-{1\over3}\,\unitvec e_z-{\sqrt2\over3}\,\unitvec e_x
\pm\sqrt{{2\over3}}\,\unitvec e_y\;.
\label{eq:tet2}
\end{equation}
It is easy to see that for this tetrahedron, $\sum_{j=1}^4
Y_{lr}(\unitvec n_j)$ vanishes when $r$ is not a multiple of 3.  For
$r=0$ this sum is proportional to $\sum_{j=1}^4 P_l((n_j)_z)=1+
3P_l(-1/3)$, and for $r=3n\ne0$, it is proportional to $\sum_{j=2}^4
P_l^{3n}((n_j)_z)=3P_l^{3n}(-1/3)$.  For $l=5$, it is easy to check
that
$P_5(-1/3)=-1/3$ and $P_5^3(-1/3)=0$, so the sum
condition~(\ref{eq:frame2b}) holds for $l=5$, and a frame function can
contain $l=5$ harmonics.

\subsubsection{Other Measurements}

We now apply our technique to other measurements with nice symmetry
properties: the five platonic solids, planar polygons of any degree,
and the uniform POVM.

Easiest is the case of the uniform POVM, i.e., a measurement whose
outcomes include every direction on the Bloch sphere.  There being
only one POVM in the restricted set, the frame conditions reduce to
the requirement that $F(\unitvec n)$ be a real, nonnegative function
that integrates to unity on the sphere.  In terms of a spherical
harmonic expansion, this fixes $c_{00}= 1/\sqrt{4\pi}$, while the
other coefficients are only restricted so as to provide real,
nonnegative function values.

Continuing in reverse order, consider the case when the allowed unit
vectors lie in a plane, forming a regular polygon with $N$ vertices.
For a POVM with unit vectors lying in the $x$-$y$ plane, it is easy to
see that $\sum_{j=1}^N Y_{lr}(\unitvec n_j)$ is zero unless $r$ is a
multiple of $N$; when $r=nN$, the sum is proportional to $P_l^{nN}(0)$,
which is zero (nonzero) if $l+nN$ is odd (even). As a consequence, the
allowed harmonics for $N$ even are $l=0$ and all odd harmonics, and the
allowed harmonics for $N$ odd are $l=0$ and the odd harmonics such that
$l\le N-2$.  Notice that the trine is the only regular polygon that
gives the quantum probability rule.

The vertices of each of the five platonic solids yield a symmetric
set of unit vectors for constructing restricted sets of POVMs.
For all five, the projectors onto the unit vectors sum to a
multiple of the 3-dimensional identity operator, implying that
$l=2$ harmonics are allowed; moreover, for all but the
tetrahedron, the unit vectors come in antipodal pairs, meaning
that all odd harmonics are allowed. For the octahedron, by
considering the sum $\sum_j Y_{ll}(\unitvec n)$ over the six unit
vectors that point along the $x$ and $-x$ axes and into the four
quadrants in the $y$-$z$ plane, one finds that all even harmonics
except $l=0,2$ are ruled out.  Similarly, for the cube, this same
sum over the tetrahedron~(\ref{eq:tet1}) and its antipodal points
rules out all even harmonics except $l=0,2$.  For the dodecahedron
and icosahedron, additional even harmonics are allowed, and
numerical investigation using {\em Mathematica\/} shows these to
be $l=4,8,14$.  The results for the platonic solids are summarized
in Table~\ref{table:platsol}.
\begin{table}[h]
\label{table:platsol}
\begin{center}
\begin{tabular}{|l|r|}  \hline
\emph{Platonic solid} & \emph{Allowed harmonics}\\ \hline
tetrahedron & 0, 1, 2, 5\\
octahedron & 0, 2, \& odds\\
cube & 0, 2, \& odds\\
dodecahedron\quad &\quad0, 2, 4, 8, 14, \& odds\\
icosahedron & 0, 2, 4, 8, 14, \& odds \\ \hline
\end{tabular}
\end{center}
\end{table}
\vspace{12pt}

\section{Conclusion}
Gleason's theorem reveals that from this ``measurement first'' point
of view, the quantum probability rule is implicit in the structure of
measurements themselves.  Use of generalized measurements greatly
simplifies the proof of Gleason's theorem, extends its applicability,
and reduces the conceptual overhead of quantum theory.  Attention is
thereby refocused to the tougher task of justifying the form of
quantum measurements~\cite{Fuchs02,foulis95,ludwig83}.

Investigating particular measurements and their allowable probability
distributions complements the investigation of Kochen-Specker
colorable sets~\cite{Cabello03}.
Considering particular sets of POVMS allows us to
explore the range of probability distributions that lie between the
quantum probability rule and Kochen-Specker truth assignments.

This work naturally leads to the question of whether POVMs ought to be
considered, in some sense, more fundamental than standard
projection-valued measures.  The reason for thinking so is not just the
simplicity of proofs.  Foremost is the notion of ``fuzziness'' that
effects capture, a notion essential for practical purposes. Secondly,
in thinking of quantum mechanics operationally, nothing singles out
projection measurements for {\em fundamental\/} status. While it's true
that effects themselves are convex combinations of projection
operators, POVMs needn't be convex combinations of projection-valued
measures~\cite{DAriano03}.  Finally, POVMs are \emph{useful}
measurements, as demonstrated by their increasing use in the field of
quantum information~\cite{Nielsen00}. All these reasons point toward a
fundamental role for POVMs.

\section*{Acknowledgments}
This work was supported in part by Office of Naval Research
Grant No.~N00014-00-1-0578.

\end{document}